\documentclass[natbib, twocolumn]{svjour3}         
\smartqed  
\usepackage{graphicx}
\usepackage{mathptmx}      
\usepackage{aps-bibstyle}  
\newcommand{\alfven}{Alfv\'{e}n}
\newcommand{\alfvenic}{Alfv\'{e}nic}
\newcommand{\sdo}{{\em SDO}}
\newcommand{\hinode}{{\em Hinode}}
\newcommand{\dof}{{$\mathrm{{D}_{X}}$}}
\newcommand{\RNum}[1]{\uppercase\expandafter{\romannumeral #1\relax}}
\newcommand\ion[2]{#1$\;${\RNum{#2}}}%

\newcommand{\pref}{\protect\ref}

\newcommand{\aap}{{Astron. Astrophys.}}
\newcommand{\aaps}{{Astron. Astrophys. Supp.}}
\newcommand{\apj}{{Astrophys. J.}}
\newcommand{\apjs}{{Astrophys. J. Supp.}}
\newcommand{\apjl}{{Astrophys. J. Lett.}}
\newcommand{\grl}{{Geophys. Res. Lett.}}
\newcommand{\jgr}{{J. Geophys. Res.}}
\newcommand{\solphys}{{Solar Phys.}}
\newcommand{\ssr}{{Space Sci. Rev.}}
\newcommand{\pasj}{{Proc. Ast. Soc. Japan}}
\newcommand{\nat}{{Nature}}

\begin{document}

\title{Recent Observations of Plasma and \alfvenic{} Wave Energy Injection at the Base of the Fast Solar Wind}
\titlerunning{The Roots of the Fast Solar Wind}

\author{Scott W. McIntosh}

\institute{Scott W. McIntosh \at
              High Altitude Observatory, 
              National Center for Atmospheric Research, \\
              P.O. Box 3000, Boulder, CO 80307 USA\\
              Tel.: +001-303-497-1544\\
              Fax: +001-303-497-1589\\
              \email{mscott@ucar.edu}}

\date{Received: date / Accepted: date}

\maketitle

\begin{abstract}
We take stock of recent observations that identify the episodic plasma heating and injection of \alfvenic{} energy at the base of fast solar wind (in coronal holes). The plasma heating is associated with the occurrence of chromospheric spicules that leave the lower solar atmosphere at speeds of order 100km/s, the hotter coronal counterpart of the spicule emits radiation characteristic of root heating that rapidly reaches temperatures of the order of 1MK. Furthermore, the same spicules and their coronal counterparts (``Propagating Coronal Disturbances''; PCD) exhibit large amplitude, high speed, \alfvenic{} (transverse) motion of sufficient energy content to accelerate the material to high speeds. We propose that these (disjointed) heating and accelerating components form a one-two punch to supply, and then accelerate, the fast solar wind. We consider some compositional constraints on this concept, extend the premise to the slow solar wind, and identify future avenues of exploration.
\end{abstract}

\section{Introduction}\label{intro}
The observational study of coronal holes permits the investigation of mass and energy transport through the lower solar atmosphere. This transport directly impacts the mass content and dynamics of the fast solar wind. The energetics of the lower solar atmosphere in a coronal hole are {\em identical} to those of the quiet Sun and studying the base of coronal holes permits an investigation into the basal mass and energy injection of the quiet corona without the energetic complications that arise from a closed magnetic topology \citep[see e.g.,][and later]{2007ApJ...654..650M,2012ApJ...749...60M}. The (often subtle) observational differences between the two ``modes'' of the quiescent solar atmosphere stem from the fact that material trapped in the quiet corona has the ability to return to the lower solar atmosphere \-- a process that (in the large part) does not happen in coronal holes. The complete ``chromosphere-corona mass cycle'' of the quiet solar atmosphere modifies the spectral characteristics of the emission observed because both the emission components due to the heating upward mass flux {\em and} cooling return flow can be observed in a single spatio-temporal resolution element \citep[e.g.,][]{2012ApJ...749...60M}. Furthermore, it is expected that the mass cycle can affect the compositional differences that are measured in solar wind streams that originate from open and closed magnetic regions, a topic we will briefly return to in the penultimate section of this paper.

The {\em SOHO} \citep[][]{1995somi.book.....F} generation of coronal hole spectroscopic studies have significantly improved our understanding of the relentless mass and energy transport at the base of the heliosphere \citep[see, e.g.,][and references therein]{1999A&A...346..285D,1999Sci...283..810H,2003A&A...399L...5X,2005Sci...308..519T,2006ApJS..165..386D,2006ApJ...644L..87M}. Recently, a picture of energy release and initial mass supply to the corona and solar wind originating in the lower atmosphere has developed following the observation of a finely scaled, high velocity component of the spicule family \citep[][]{2007PASJ...59S.655D}. These ``Type-II'' spicules have been associated with a weak but ubiquitous signature of chromospheric mass supply to the outer solar atmosphere rooted in unipolar magnetic field regions \citep[through the study of transition region and coronal spectroscopy, e.g.,][]{2009ApJ...701L...1D,2009ApJ...707..524M,2010A&A...521A..51P,2010ApJ...722.1013D,2011ApJ...727L..37T} and broadband coronal imaging sequences \citep[e.g.,][]{2009ApJ...706L..80M, 2011ApJ...736..130T}. The detailed correspondence of dynamic chromospheric and coronal plasma heating has recently been confirmed through joint observation from {\em Hinode} \citep[][]{2007SoPh..243....3K} and the {\em Solar Dynamics Observatory} (\sdo) that are presented by \citet{2011Sci...331...55D}. In that paper discrete high-velocity heating jets triggered in the lower-atmosphere have been uniquely traced through the chromosphere, transition region, and into the outer atmosphere (as ``Propagating Coronal Disturbances'' or PCDs) for the first time.

Another recent observational development relevant to the basal energetics of the fast solar wind has been the identification of a sufficient \alfvenic{}\footnote{The interested reader is pointed to the supporting material of \citet{2011Natur.475..477M} where a lengthy discussion defines their use of this term to describe the observed motion, relative to the ``kink'' or ``\alfven'' which require a more precise description of the plasma and its magnetic field than can be determined observationally.} wave energy flux to propel the fast wind. \citet{2007Sci...318.1574D} inferred that an abundant \alfvenic{} wave flux was present in the chromosphere, visible in the motion of {\em all} chromospheric spicules, but was particularly clear to observe in the polar coronal holes that are dominated by Type-II spicules. Due to the lack of coronal observations with commensurate spatio-temporal resolution with the \hinode{} Solar Optical Telescope, it was unclear at that time wether or not the observed wave flux was reflected at the chromosphere-coronal boundary. A subsequent observational study by \citet{2011Natur.475..477M} used \sdo{} observations to infer that the wave amplitude and phase speed throughout the solar atmosphere through the motions of transition region spicules and PCDs at coronal temperatures. The observed 100-200Wm$^{-2}$ of wave energy in coronal holes resided in \alfvenic{} motions of 300-500s period, 25km/s amplitude and 1Mm/s phase speed --- a flux of non-WKB waves streaming outward on structures that weakly emit at temperatures of order 1MK\footnote{The emission in which the PCDs are most clearly observed is formed (in equilibrium) at $\sim$1MK. It is unclear given the rapid nature of the heating readily visible at the base of these structures (see below) wether or not the plasma is in ionization equilibrium or not in during the initial heating phase. It is likely that some of the material is in excess of 1MK upon breaching the Sun's atmosphere.}. This wave flux is sufficient to drive the fast solar wind \citep[e.g.,][]{1995JGR...10021577H}.

Putting the observational results of \citet{2011Sci...331...55D} and \citet{2011Natur.475..477M} together, we have two components at work in the magnetic elements that comprise the supergranular network \--- the tributaries of the fast solar wind \--- (quasi-periodic) mass heating and injection to temperatures of order 1MK, and a significant flux of long period \alfvenic{} motions that are needed to accelerate that mass away from the Sun. We take the stance that somehow, this two-stage ``engine'' works to produce what we observe in situ \citep[a concept proposed in][]{1991ApJ...372..719P}. The precise synthesis of the physical processes taking place near the inner boundary of the heliosphere that produce the fast solar wind are to be determined, but such a synthesis will require theoretical endeavor that incorporates self-consistency of the mass and wave transport in the numerical simulations \citep[e.g.,][]{2005ApJS..156..265C,2007ApJS..171..520C,2007Sci...318.1574D,2010ApJ...708L.116V}. It is the view of the author that the fine details of the fast wind initiation and acceleration mechanisms are within our reach, and that the solution of the puzzle will yield invaluable insight into the complex evolution of the slow solar wind. Unfortunately, it is also the view of the author that, in order to understand the slow solar wind, we {\em must} consider the circulation of material between the chromosphere and corona and other factors {\em before} we consider the transition of the material onto an open field line and outward into interplanetary space.

The following sections take a look at the recent observational investigations of coronal holes. We briefly discuss the importance of understanding compositional measurements of the fast solar wind in the context of the mass and energy transport going on at its roots before discussing those observational properties that must be reproduced by numerical models to encapsulate the full physics of the fast solar wind.

\section{Observational Investigations}\label{obs}
Observational investigations of coronal holes started as soon as we could routinely observe the coronal plasma from above the Earth's atmosphere \citep[e.g.,][]{1967SoPh....1..129U,1968Sci...162...95G,1968IAUS...35..395B,1972ApJ...176..511M,1974ApJ...194L.115H}. Soon after \citet{1973SoPh...29..505K} realized that the large dark patches in the corona were spatially related (and physically connected) to high speed solar wind streams \--- coronal holes soon became a topic of hot discussion (pardon the pun). These initial investigations are fascinating and highlight the apparent over-abundance of iron in the quiet corona relative to the photosphere \citep[][]{1968Sci...162...95G}, a clear pre-cursor to the ``FIP Effect" \citep[e.g.,][]{1995Sci...268.1033G}. In the following subsections we will present the author's view of coronal holes, considering SOHO/SUMER \citep[][]{1995SoPh..162..189W} spectroscopic and SDO/AIA imaging observations to infer the processes occurring at the roots of the fast solar wind.

\subsection{Spectroscopic Investigations}\label{spec}
\begin{figure}
\begin{center}
\includegraphics[width=85mm]{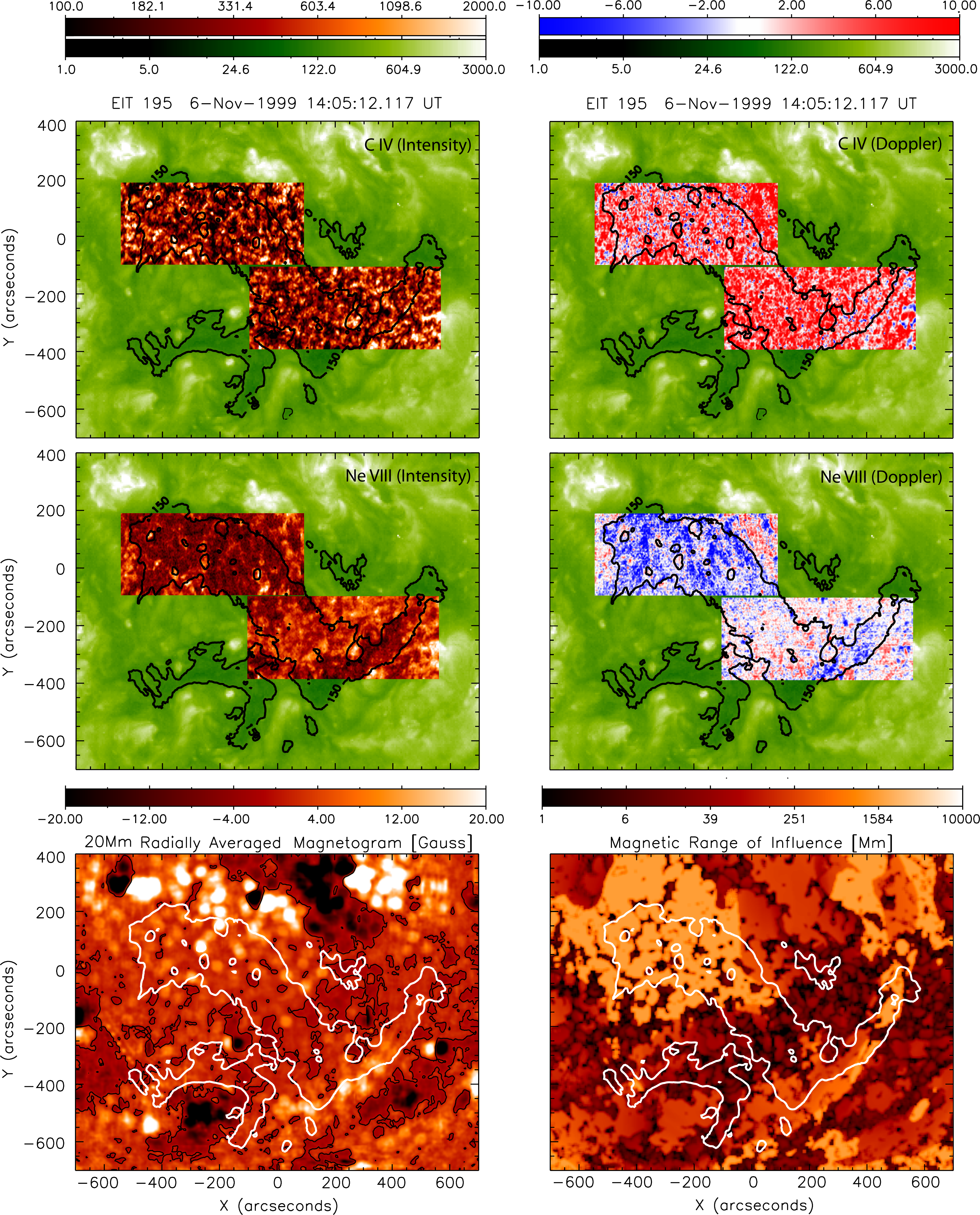}
\caption{\citet{2006ApJ...644L..87M} and \citet{2007ApJ...654..650M} investigated the spectroscopic structure of an ECH observed by SUMER in early November 1999. This figure shows the comparison of two simple magnetic diagnostics and their correlation to the \ion{C}{4} and \ion{Ne}{8} intensity and Doppler velocity maps. The top two rows show the SUMER \ion{C}{4} 1548~\AA{} and \ion{Ne}{8} 770~\AA{} line intensities (left) and Doppler-velocities (right) respectively. The bottom row shows the supergranular radially averaged MDI magnetic field strength and the ``Magnetic Range of Influence'' \citep[MRoI; ][]{2006ApJ...644L..87M}. The MRoI is a distance measure of balance in the magnetic field, e.g., in regions where there is local unipolarity in the field, the distance to find enough opposite polarity magnetic flux is large. On each of the panels in the figure, we show the EIT 195~\AA{} 150~DN contour to outline the ECH. Additionally, in the lower left panel, we show the thin black contour that designated the magnetic ``neutral line'', where B=0~G. Notice the very strong correspondence between regions of large MRoI and strong \ion{Ne}{8} blue shift.}\label{f1} 
\end{center}
\end{figure}

Observations from the SOHO SUMER, EIT \citep[][]{1995SoPh..162..291D} and MDI \citep[][]{1995SoPh..162..129S} instruments have provided  breakthrough insight into the origins of the fast solar wind in coronal holes \citep[e.g.,][]{1999A&A...346..285D,1999Sci...283..810H,2003A&A...399L...5X,2005Sci...308..519T}. SUMER ``spectroheliogram'', or raster, observations were employed to quantitatively examine the origins of the solar wind by correlating Doppler velocity measurements of \ion{Ne}{8} (formed at about 600,000K in the upper solar transition region) to a proxy for the gross super-granulation pattern of chromospheric network structure. Results of these prior studies suggested a clear correlation between the observed \ion{Ne}{8} blue shift\footnote{The convention being that a blue shift is associated with a negative velocity and as such a blue Doppler shift indicates the presence of plasma moving towards the observer, possibly indicating an out-flow from the Sun.} and the chromospheric network pattern. Unfortunately, many of these earlier investigations focused on polar coronal holes where line-of-sight (LOS) plasma diagnostics are not always ideal. To address this issue, \citet{2006ApJ...644L..87M} studied an equatorial coronal hole (ECH) with SUMER to minimize the impact of high-latitude observations on the LOS spectroscopy. Indeed, they demonstrated that while there is indeed a strong correlation between bright network emission and \ion{Ne}{8} blue Doppler-shifts, there is a substantial fraction of blue shifted \ion{Ne}{8} plasma in coronal holes that is not correlated with underlying network emission of the supergranular boundaries (or bright network vertices), see Fig.~\pref{f1}. This analysis would suggest that smaller scale upflows exist throughout the ECH and probably throughout other coronal holes, as they all have the same intrinsic property of a local polarity imbalance of the magnetic field (see e.g., the bottom figures of Fig.~\pref{f1}). 


\begin{figure}
\begin{center}
\includegraphics[width=85mm]{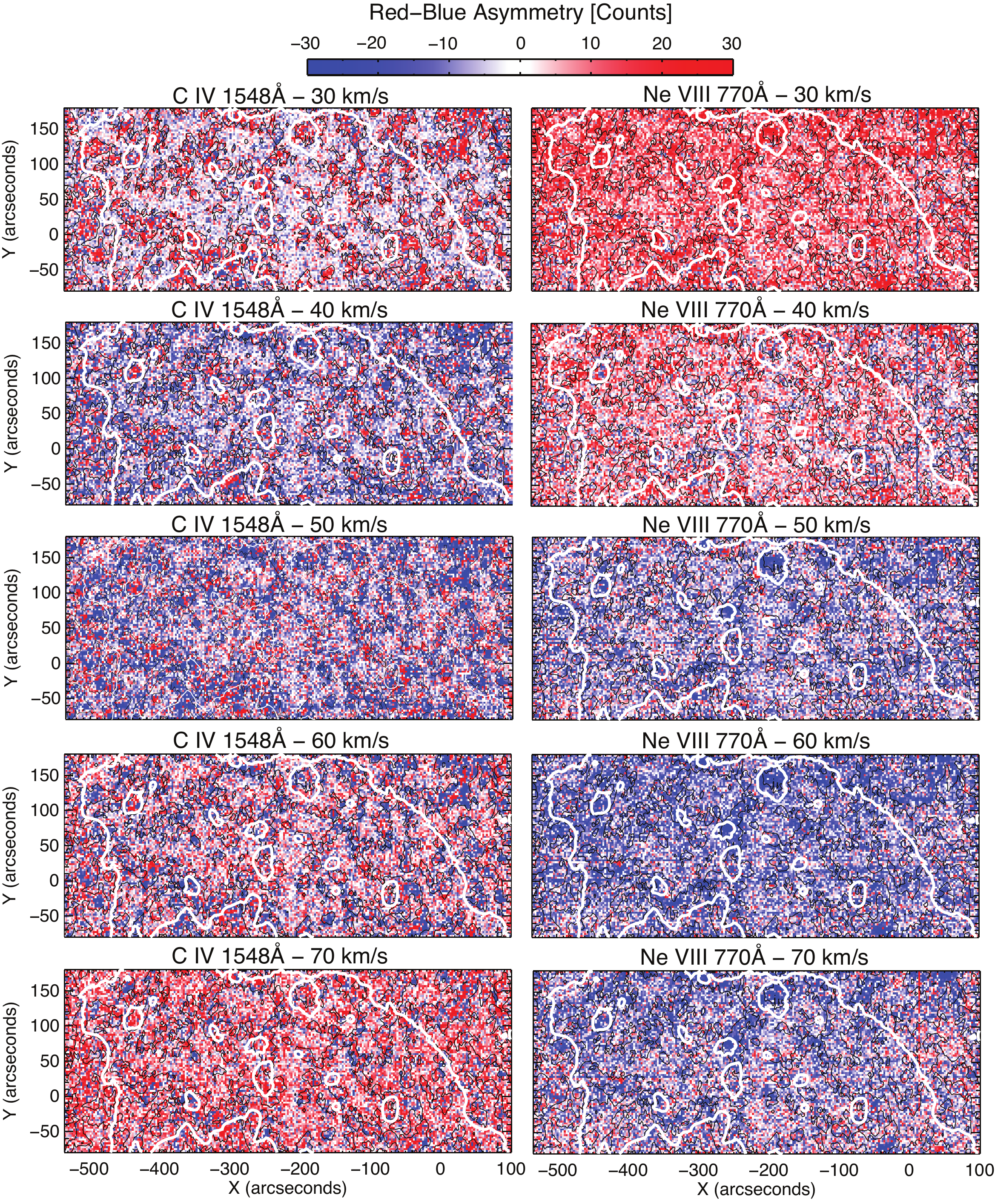}
\caption{Adapted from \citet{2011ApJ...727....7M}. The velocity dependence (top to bottom) of the \ion{C}{4} (left) and \ion{Ne}{8} (right) ``R-B'' line profile asymmetry analysis for the upper left portion of the ECH shown in Fig.~\pref{f1}. The white contours show the outline of the ECH and the finer black contours show the locations of strong network emission. Blue wing asymmetries are prevalent in the magnetic network of the supergranular vertices.}\label{f2} 
\end{center}
\end{figure}

Shortly following the discovery of high velocity Type-II spicules with \hinode{} SOT in coronal holes, we started an investigation to identity if the rapid fading and upward motion of these very dynamic events were a marker of intense plasma heating in the lower atmosphere. Further, we were interested to see if they had an observable signature in hotter emission that would indicate a possible role in mass transport to the outer solar atmosphere. \citet{2009ApJ...701L...1D} inferred a connection between the apparent motion of Type-II spicules and a weakly emitting component of emission observed far in the blue wing of coronal emission lines formed at a broad range of temperatures, reaching at least 2MK \citep[as observed by \hinode{} EIS][]{2007SoPh..243...19C}\footnote{This artificial upper bound is driven by limited observation of relatively ``clean'' spectroscopic emission lines in the wavelength range studied by EIS.}. This subtle signature (at 5-15\% of the background emission) was only visible in the highest signal-to-noise spectroscopic datasets from \hinode{}, but it appeared to have a very strong effect on the observed ``non-thermal'' broadening of the line profiles over magnetized regions \citep[e.g.,][]{2008ApJ...678L..67H}. \citet{2009ApJ...701L...1D} developed a diagnostic measure of additional emission components present in spectroheliograms which differences the amount of emission in the blue and red wing of a line profile. The difference is interpreted as a net imbalance of upward (blue) or downward (red) emission at that velocity (relative to the measured center of the line profile). This Red-Blue (or ``R-B'') diagnostic was subsequently applied to SUMER quiet Sun \citep[QS;][]{2009ApJ...707..524M} and ECH \citep[see Fig.~\pref{f2} and][]{2011ApJ...727....7M} observations, where a residual high speed component of the emission was observed to coincide with the magnetized network locations. This indicates that the weakly emitting high-speed upflow events reached at least temperatures comparable with the (equilibrium) formation temperature of \ion{Ne}{8} in coronal holes. We also note that the regions with enhanced non-thermal broadening in the ECH are intimately related to the locations where the R-B diagnostic indicates blue-wing asymmetric line profiles. One possible interpretation of the enhanced non-thermal broadening historically observed is that in magnetized regions of coronal holes is that more than one spectral component contributing to the measured emission line profile. The supposition is that these upflow events are triggered by (component) magnetic reconnection in the unipolar supergranular network vertices, and that the material is  heated and ejected outward from the Sun's surface in the process \citep[e.g.,][]{2011ApJ...736....9M}.

Significant limitations face the R-B analysis for archived measurements, such as those from SUMER: including the typically low S/N of the line spectra significantly impacts our ability to diagnose a high velocity component with 5-10\% of the core intensity. Also, spectral blends of known (and unidentified) emission lines can severely contaminate the diagnosis. However, the trends observed by R-B analyses using many different spectral lines in different wavelength ranges, and for a range of formation temperatures \citep[e.g.,][]{2009ApJ...707..524M, 2011ApJ...732...84M}, support the assertion that spectral blends are {\em not} the cause of the systematic 50-100km/s blue wing asymmetries observed in magnetized regions\footnote{Higher velocities are typically observed above plage regions than the QS/CH network.}. It would be reasonable to assume, therefore, that the R-B measure is indeed detecting the signature of weakly emitting, rapidly upflowing, material moving at approximately the same speed for different temperatures in the magnetized regions at the base of the fast solar wind. Unfortunately, the very low sensitivity of \hinode{} EIS to lines formed at temperatures below $\sim$1MK has rendered ECH studies with that spectrograph problematic. The upcoming Interface Region Imaging Spectrograph (IRIS) mission we will be able to explore the complex line profiles in emission lines spanning the chromosphere, transition region, and lower corona with very high S/N, high spatio/temporal/spectral resolution, and in a set of unblended emission lines such that many of the R-B diagnostics can be directly verified.

\begin{figure}
\begin{center}
\includegraphics[width=65mm]{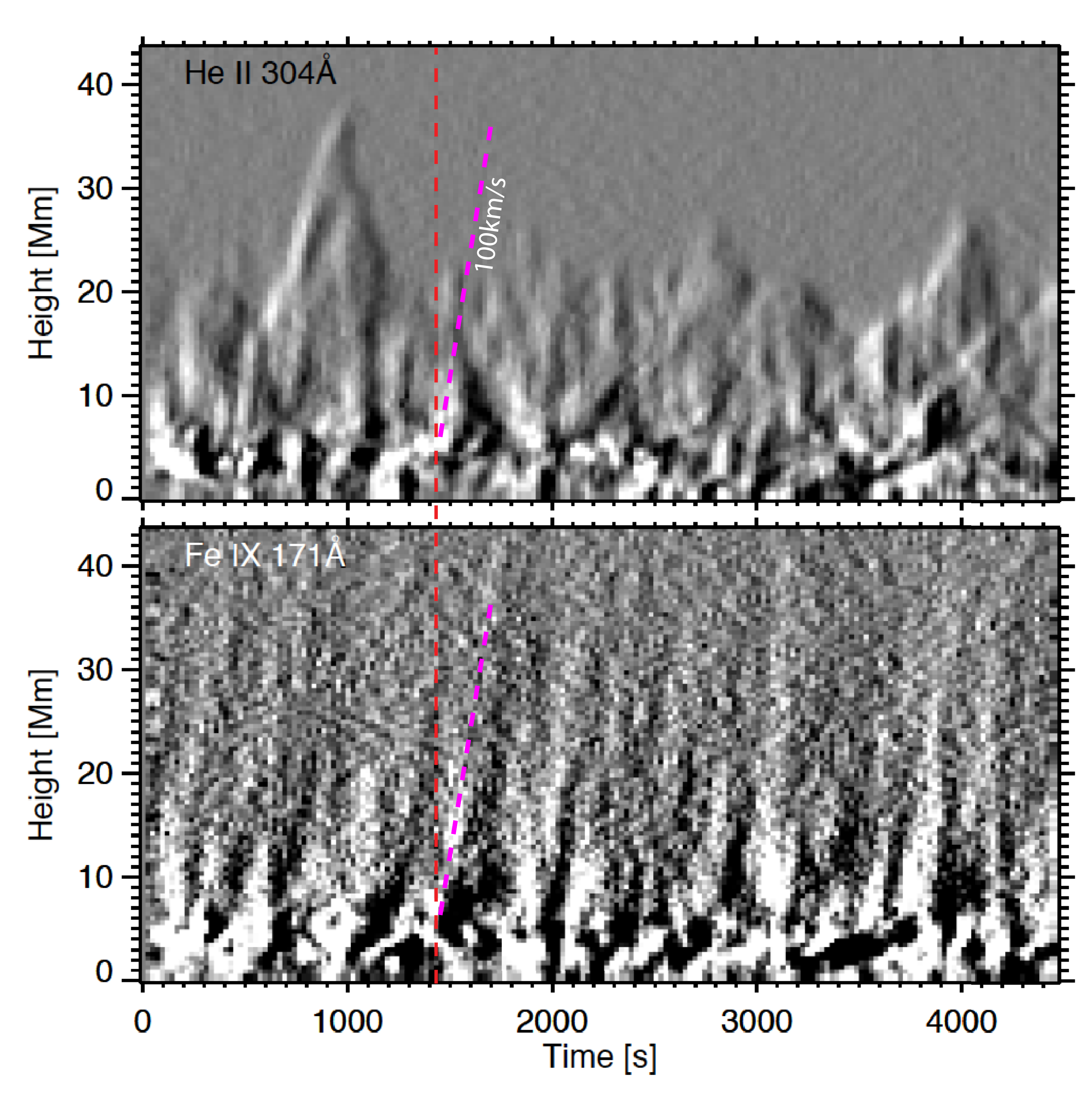}
\caption{Adapted from \citet{2011Sci...331...55D}. The temporal and thermal evolution of off-limb spicules in a coronal as observed in the SDO/AIA \ion{He}{2} 304\AA{} (top) and \ion{Fe}{9} 171\AA{} (bottom) passbands, summing three images in time to boost the visibility of the faint off-limb signals, with a resulting cadence of 24s. Calculating space-time plots for a ``cut'' perpendicular to the limb, i.e., parallel to the prevailing spicules and the resulting space-time plots of the running difference (over 3 timesteps, i.e., 72s) show many clear parabolic (up and down) paths in \ion{He}{2}  304\AA{} that are often associated with upward propagating disturbances in \ion{Fe}{9} 171\AA{}  (PCDs) \-- the red vertical dashed line highlights one spicule/PCD correspondence. The \ion{He}{2} 304\AA{} spicules have lifetimes of order several minutes. The apparent propagation speed of the PCD and initial phase of the spicules are of order 100 km/s (the pink dashed line is shown as a reference).}\label{f3} 
\end{center}
\end{figure}

\subsection{Coronal Imaging Investigations}\label{image} 
The high S/N, low scattered light of the SDO/AIA telescopes permit the detailed imaging of the high velocity upflow events that were inferred from the SUMER spectroscopic measurements in the quiet sun and coronal holes. These imaging investigations follow from the analysis of \citet{2007Sci...318.1585S} who identified episodic, high speed flows on coronal loop structures in lower spatial resolution \hinode{} XRT observations that were rooted in the underlying strong magnetic field regions \citep[also see][for details of the correspondence between these events and the spectroscopic observations of line profile asymetries]{2009ApJ...707..524M, 2011Sci...331...55D, 2011ApJ...738...18T}.

\citet{2011Sci...331...55D} demonstrated that the on-disk counterpart of Type-II spicules had an associated coronal signature \--- a propagating coronal disturbance, or PCD. They interpreted the observations as an indication that there was strong plasma heating associated with the formation of the spicule \--- in active regions, the associated feature was visible in emission characteristic of plasma at 2~MK\footnote{Unfortunately the S/N in the ``hotter'' AIA passbands is significantly lower, and so it is unclear what the maximum temperature of the material in the PCD is.}. The supporting online material of that paper used limb observations in a polar coronal hole to illustrate the correspondence between the lower atmosphere and the corona (Fig.~\pref{f3}). Even with the significantly better observing conditions of SDO/AIA, establishing a one-to-one connection is difficult due to the massive LOS of the optically thin emission at the limb. Space-time (X-T) plots are constructed by monitoring the evolution of the plasma in the first few Mm perpendicular to the limb in the \ion{He}{2} 304\AA{} (top) and \ion{Fe}{9} 171\AA{} (bottom). We see that there are many instances where the initial phase of the transition region spicule (see the red dashed vertical line as an example) is shared with the launch of the PCD, indeed both features have the same initial speed (the pink dashed line indicates an apparent speed of 100km/s). Like in active regions, the observations of \citet{2011Sci...331...55D} deduced that the plasma of the lower solar atmosphere in the magnetized regions of coronal holes underwent rapid upward motion and heating.

\begin{figure}
\begin{center}
\includegraphics[width=85mm]{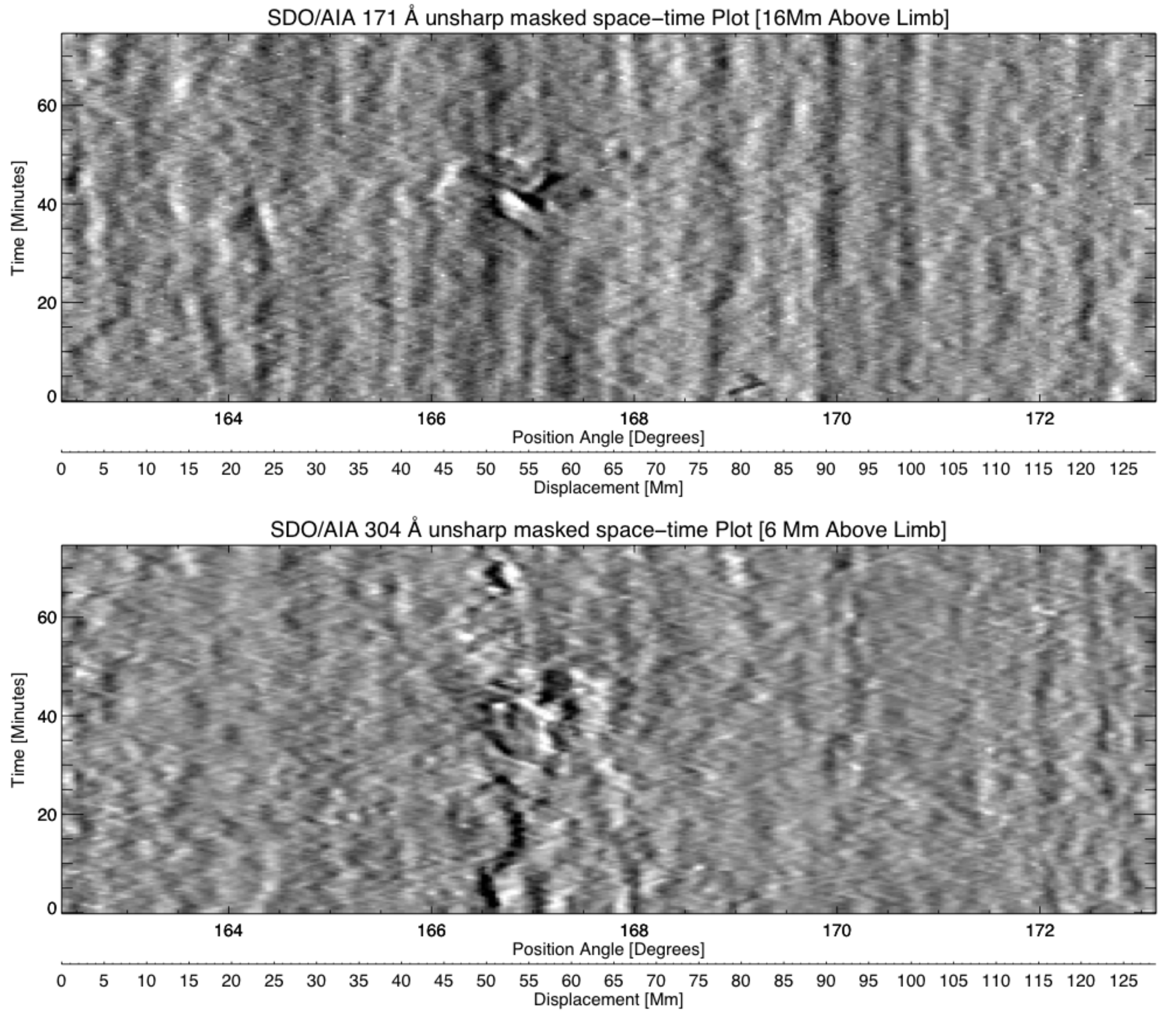}
\caption{Adapted from the supporting online material of \citet{2011Natur.475..477M}. The transverse motion of transition region spicules (bottom) and associated PCDs (top) derived from in space-time plots taken parallel to the solar limb of a polar coronal hole (cf. Fig.~\pref{f3}). The \alfvenic{} motions are rarely complete sinusoids because of the short spicule/PCD lifetimes. We also see wave motions that are horizontally separated by as much as 5~Mm that have in-phase transverse oscillations. This provides support for our assertion that the waves are volume filling.}\label{f7} 
\end{center}
\end{figure}

\subsubsection{The Transverse (\alfvenic) Motion of PCDs}\label{image} 
As we have seen above, the AIA \ion{He}{2} 304\AA{} channel at the solar limb shows a transition region that is dominated by spicular jets that shoot rapidly upwards, and often those jets reach heights of 20,000km above the solar limb (Fig.~\pref{f3}). Observations of the same region in the \ion{Fe}{9} 171\AA{} channel reveal associated PCDs that propagate outward at high speeds ($\sim$100 km/s). When studying space-time plots parallel to the solar limb it is clear that these transition region and coronal features undergo significant \alfvenic{} motion with displacements varying sinusoidally in time \citep[see, e.g., Fig.~1 of][]{2011Natur.475..477M}. Indeed, SDO/AIA image sequences of polar coronal holes show an outer atmosphere that is replete with \alfvenic{} motion. The waves are traced by structures that do not have long lifetimes (of order 50-500s) compared to the wave periods ($\sim$5~minutes), and are difficult to detect because of the enormous LOS superposition above the solar limb. These factors contribute to the fact that very few complete swings of the spicule (or PCD) are observed and we are left with the ``criss-cross'' pattern of temporal evolution at a specific height above the limb (Fig.~\pref{f7}). \citet{2011Natur.475..477M} followed the analysis of \citet{2007Sci...318.1574D} and used Monte Carlo simulations to study the patterns produced by the propagation of the of the transverse motion. They found that the coronal hole waves had periods in the range of 150-550s, and amplitudes of order 25km/s in emission characteristic of coronal temperatures, i.e. clearly some portion of the \alfvenic{} energy had made it ``into the corona'' without being reflected. Using cross-correlation techniques of parallel space-time plots at different heights above the polar limb to determine the phase speed of the \alfvenic{} motions (reaching 1~Mm/s at an altitude of 50Mm) \citet{2011Natur.475..477M} identified that the volume filling waves carried somewhere between 100 and 200 Wm$^{-2}$ at the lower boundary of the fast solar wind.

\begin{figure}
\begin{center}
\includegraphics[width=85mm]{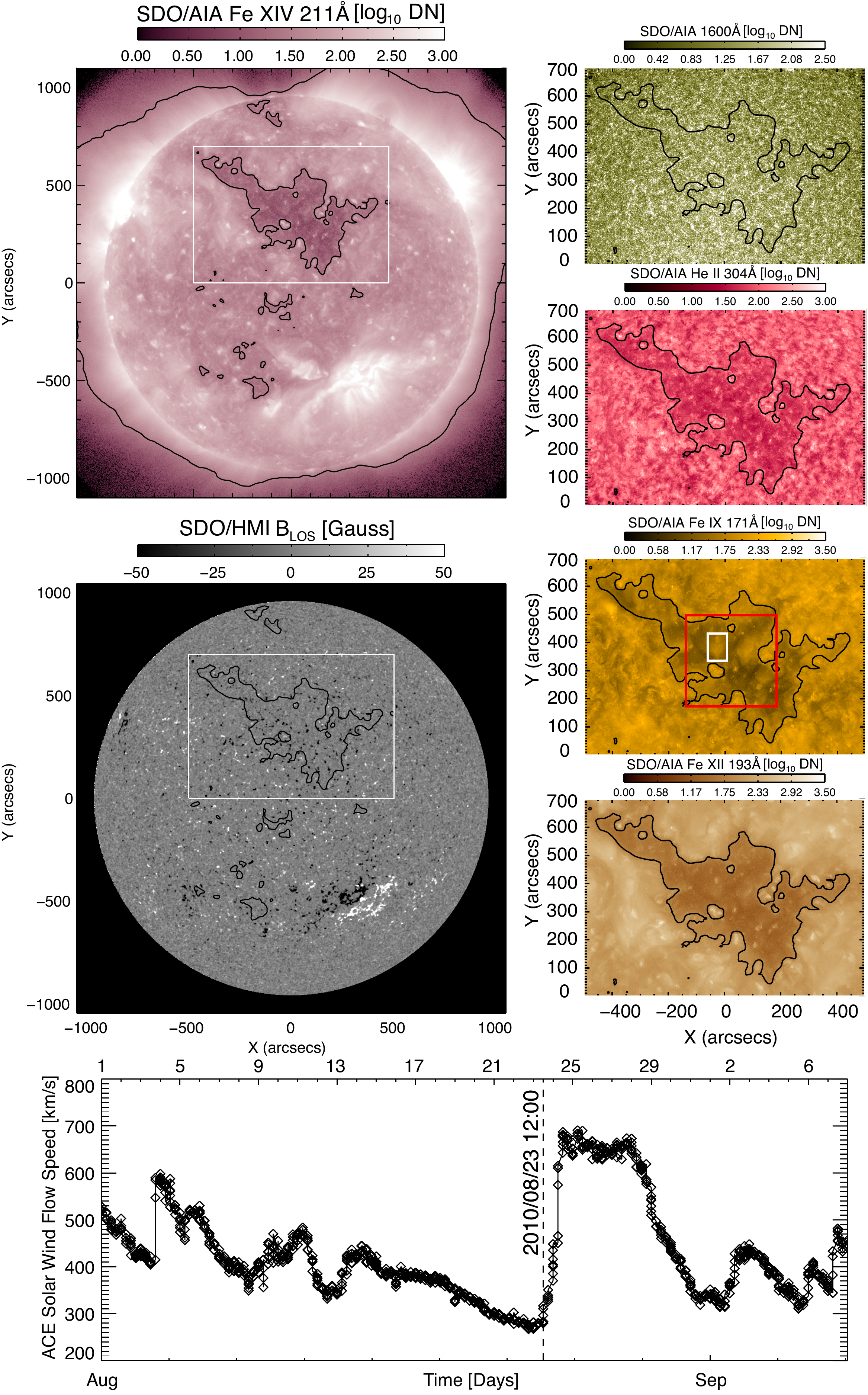}
\caption{Observational context for the ECH observations of 2010 August 23. The upper panels show the imaging data from SDO (images nearest to 12:00UT), while the lowest panel shows the hourly-averaged solar wind speed measurements from ACE/SWEPAM from 2010 August 1 to 2010 September 8. The left column shows the \ion{Fe}{14} 211\AA{} SDO/AIA channel image showing a 30~DN contour level (representing the coronal hole boundary, which is also drawn in the other panels of this figure), and the SDO/HMI line-of-sight magnetogram. From top to bottom, the right column shows the SDO/AIA images in the cut-out region of the ECH, the 1600\AA{} UV continuum channel, \ion{He}{2} 304\AA{}, \ion{Fe}{9} 171\AA{}, \ion{Fe}{12} 193\AA{}. The electronic edition of the Journal provides individual movies of the SDO/AIA channels over the two hours of observation. The small white box in the 171\AA{} channel image is the region selected for analysis in Fig.~\pref{f5}, and the red square gives the reader an opportunity to study the fine scale activity visible throughout the ECH.}\label{f4} 
\end{center}
\end{figure}

\subsubsection{ECH Observations with SDO/AIA}
As an extension of this imaging effort we consider a large ECH presents a source of fast solar wind at 1AU as measured by {\em ACE} (see Fig.~\pref{f4}). The \sdo{} observations studied were taken on August 23 2010 from 12:00\--14:00UT in the 1600\AA{} UV continuum, \ion{He}{2} 304\AA{}, \ion{Fe}{9} 171\AA{}, \ion{Fe}{12} 193\AA, and \ion{Fe}{14} 211\AA{} channels. The cadence of the images in all of the channels was 12 seconds, except for 1600\AA{} where it was 24 seconds. Figure~\pref{f4} shows context for the observations where we use the \ion{Fe}{14} 211\AA{} image (top left) to identify the ECH boundary (a 30~Data Number contour is drawn on the image), and the HMI line-of-sight magnetogram at 12:00UT (middle left) is used for context of the underlying magnetism. We extract a 1000" x 700" rectangular region around the ECH for further analysis, and the images from the other channels are shown, from top to bottom, in order of (equilibrium) formation temperature. The image sequence for each AIA channel was then de-rotated and co-aligned using standard cross-correlation techniques to form an image cube. Then, to increase the S/N in each cube, we co-add three frames such that the effective cadence in each of the five EUV channels is 36 seconds. In the 171\AA{} panel we show a small rectangular region in the body of the ECH for detailed analysis - the online edition of the Journal carries supporting movies of each of the AIA channels over the time studied. In addition, the online edition of the Journal also contains a 171\AA{} movie that is zoomed-in to highlight finer scale structure and evolution in the ECH (red square region).

\begin{figure}
\includegraphics[width=85mm]{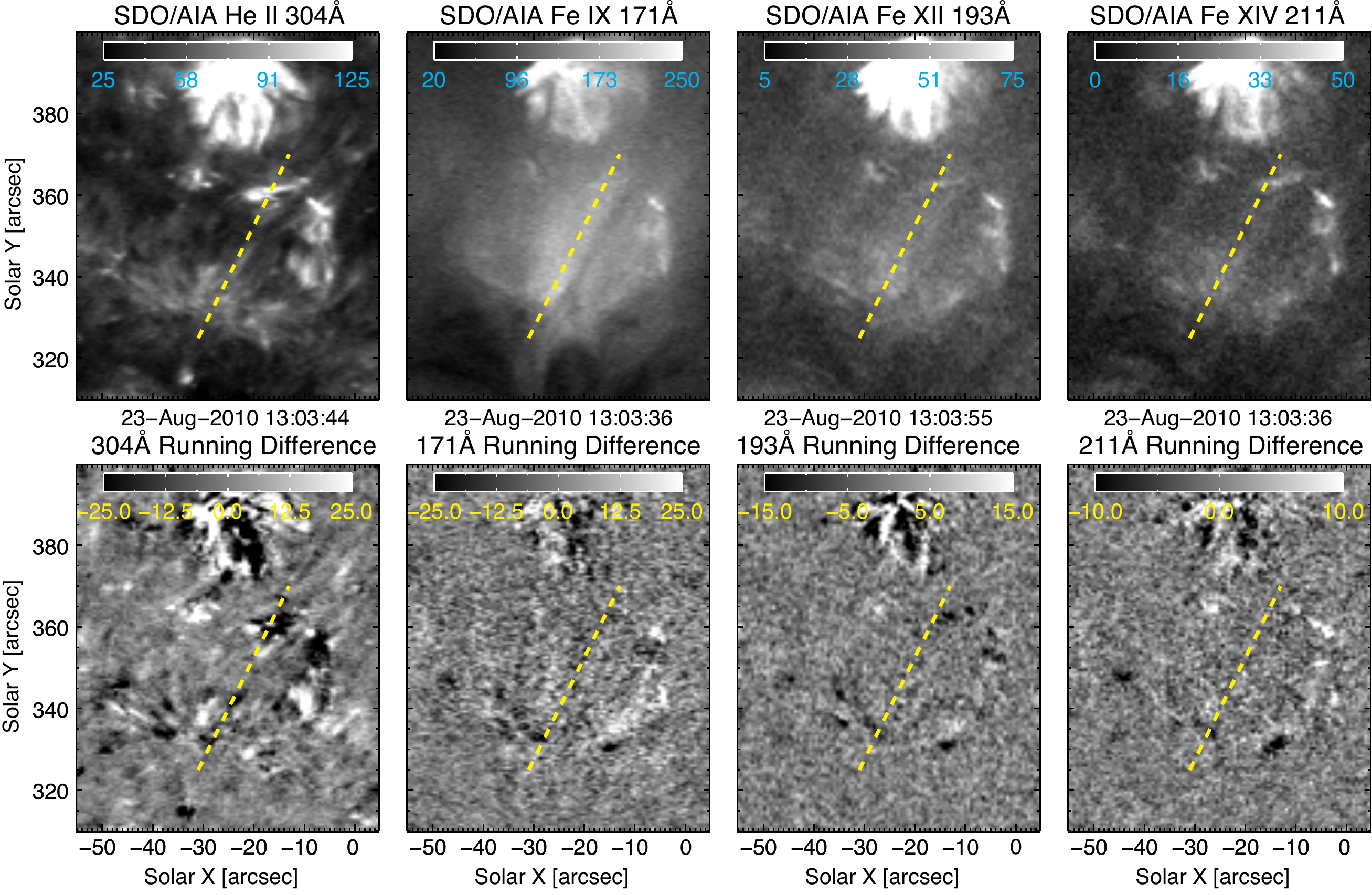}
\caption{Cut-out of a typical ECH network element. The top row of panels, from left to right, show the region in the \ion{He}{2} 304\AA{}, \ion{Fe}{9} 171\AA{}, \ion{Fe}{12} 193\AA{}, and \ion{Fe}{14} 211\AA{} channels. The bottom row of panels show the corresponding running time difference between the current frame and that taken 64s prior. The inclined dashed line is used to illustrate the apparent motion of the mass-motions observed in the extension of the network element (see, e.g., Fig.~\pref{f6}). The electronic edition of the journal has a movie of this figure.\label{f5}}
\end{figure}

Figure~\pref{f5} is designed to highlight the behavior of a typical supergranular network vertex in the ECH. The top row of panels show, from left to right, snapshots of that region in the 304\AA{}, 171\AA{}, 193\AA{}, and 211\AA{} channels. We see a general correspondence in the structure present in each panel although the 304\AA{}, having a much smaller scale height than the coronal emission, is considerably finer scaled. The bottom row of panels show the two frame (64 second) running difference of the same channels. This time-stepping is chosen to accentuate the variations of the plasma over that timescale and because the typical Type-II spicule lifetime is of order 100s. For later reference, we draw an inclined dashed line along a bright 171\AA{} feature rooted in the network. Examination of the online movie supporting this figure highlights the highly correlated behavior throughout this network element across the temperature range. There is a general consistency in PCD frequency, orientation, and enhancement over the background emission. In the hotter emission channels (193\AA{} and 211\AA{}) the excessive brightening occurs at the bottom of the network feature, and less frequent propagating jets are seen. This again would support the result of \citet{2011Sci...331...55D} that the spicule features originating in the upper chromosphere (or transition region) are rapidly heated to temperatures (often) in excess of 1MK in the network of the coronal hole.

\begin{figure}
\begin{center}
\includegraphics[width=85mm]{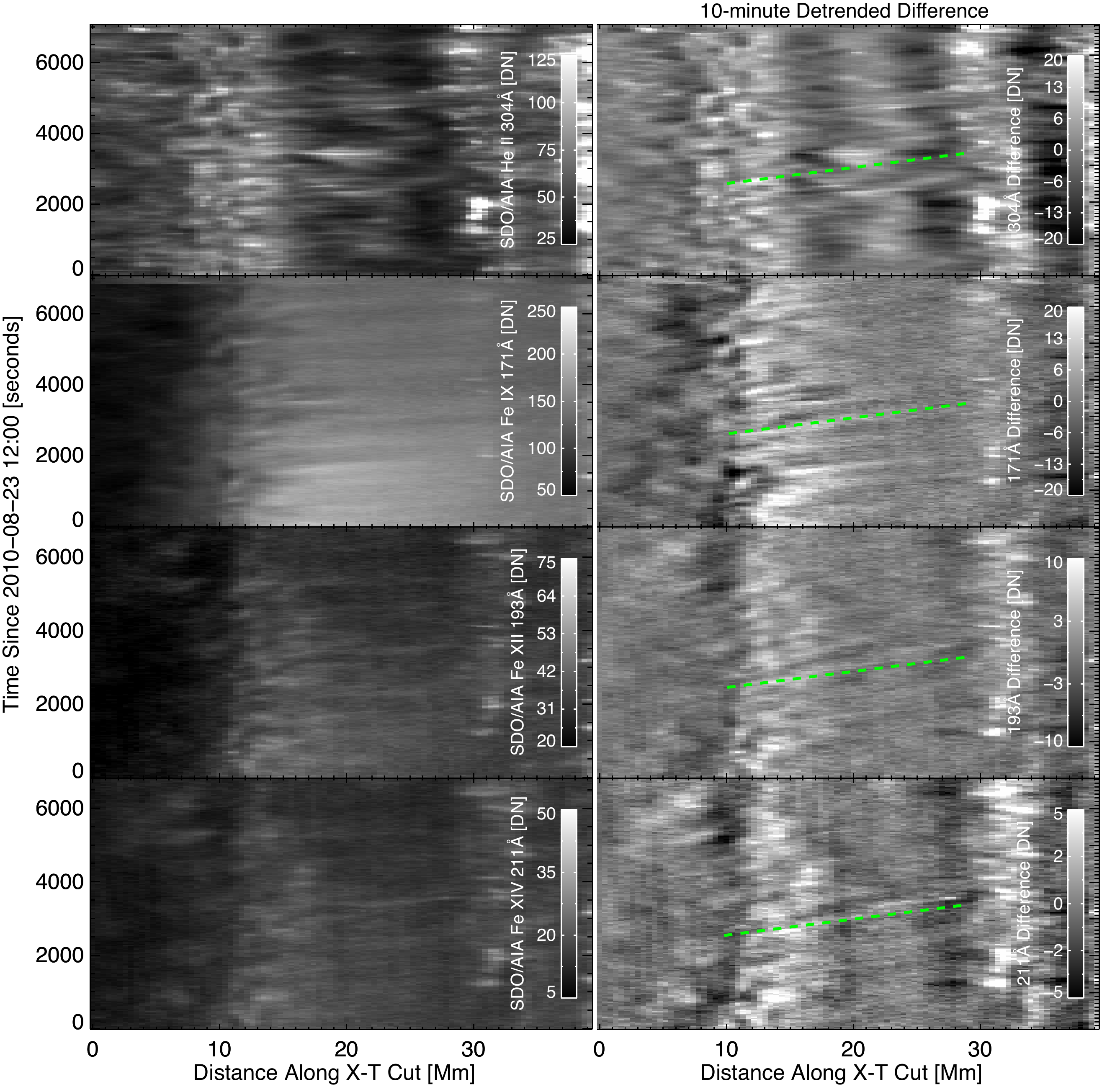}
\caption{Top to bottom, the temporal evolution of the emission (left column) and running difference (right column) in the \ion{He}{2} 304\AA{}, \ion{Fe}{9} 171\AA{}, \ion{Fe}{12} 193\AA{}, and \ion{Fe}{14} 211\AA{} channels along the dashed trajectory shown in Fig.~\pref{f5}. The dashed green line shows an apparent PCD motion of 75km/s. Using the \ion{Fe}{9} 171\AA{} channel as a reference, there are least 10 strong PCDs passing along this path in the two hours of observation.}\label{f6} 
\end{center}
\end{figure}

Using the dashed line in Fig.~\pref{f5} as a guide, we form space-time plots (shown in Fig.~\pref{f6}) to study the characteristic behavior of the plasma with time in the image sequence (left) and the image sequence with a ten-minute running average removed (right) with the panels running cool to hot from top to bottom. Throughout the timeseries we see many PCDs running along the chosen cut at {\em all temperatures} \--- the 171\AA{} channel shows at least 10 PCDs over the two hour duration of the sequence. The PCDs appear to get smaller in strength with increasing temperature, although varying degrees of S/N cannot be discounted in this effect. We see that the PCDs last on the order of 100s and that the time elapsed between consecutive PCDs can vary from $\sim$200 to $\sim$1000s. For reference, the green dashed-inclined line in the space-time diagrams indicates an apparent motion of the emission of 75km/s. The majority of the PCDs observed along this trajectory show the same apparent speed over the course of the timeseries (two hours) and across all observed temperatures.

Therefore, based on the observations present in this section, there appears to be significant plasma heating and \alfvenic{} energy supply visible in observations of the fast solar wind's lower boundary. The energy and mass release are likely driven by magneto-convective forcing of the small magnetic elements that comprise the network elements, but since they are rooted in a medium that is relentlessly undergoing convective motion the passage of the \alfvenic{} motions along the same structure would be a natural consequence. Such observations provide a means to constrain the outward mass flux and spectrum of the \alfvenic{} waves at the base of the system \--- in time measurements such as these could become inputs to numerical modeling schemes.

\subsection{Composition Measurements as Diagnostics of FSW Footprint Heating}\label{comp}
Additional evidence of the plasma processes taking place at the base of the fast solar wind are represented in the compositional measurements made in situ by the {\em ACE} or {\em Ulysses} SWICS experiments \citep[][]{1992A&AS...92..267G,1998SSRv...86..497G}. We make the assumption that the emission and rapid outward motion of PCDs at a range of high temperatures are signatures of the relentless (aggressive) plasma heating taking place in the lower atmosphere. So, the amount of energy deposited per event, and the rate at which those events occur, {\em must} be directly represented in the compositional measurements made \citep[in agreement with the assertion of][]{1995Sci...268.1033G}. If the plasma is heated to temperatures in excess of 1~MK \citep[][]{2011Sci...331...55D}, then the thermal pressure of that material is sufficient to escape the Sun's gravitational field,and the plasma will be collisionless rapidly after the heating process and insertion \citep[][]{1991ApJ...372..719P}. Thus, the mixture of ionic charge states and relative atomic abundances measured in situ in the fast wind present us with the ideal opportunity to remotely diagnose the physics of the rapid plasma heating process.

\begin{figure}
\begin{center}
\includegraphics[width=85mm]{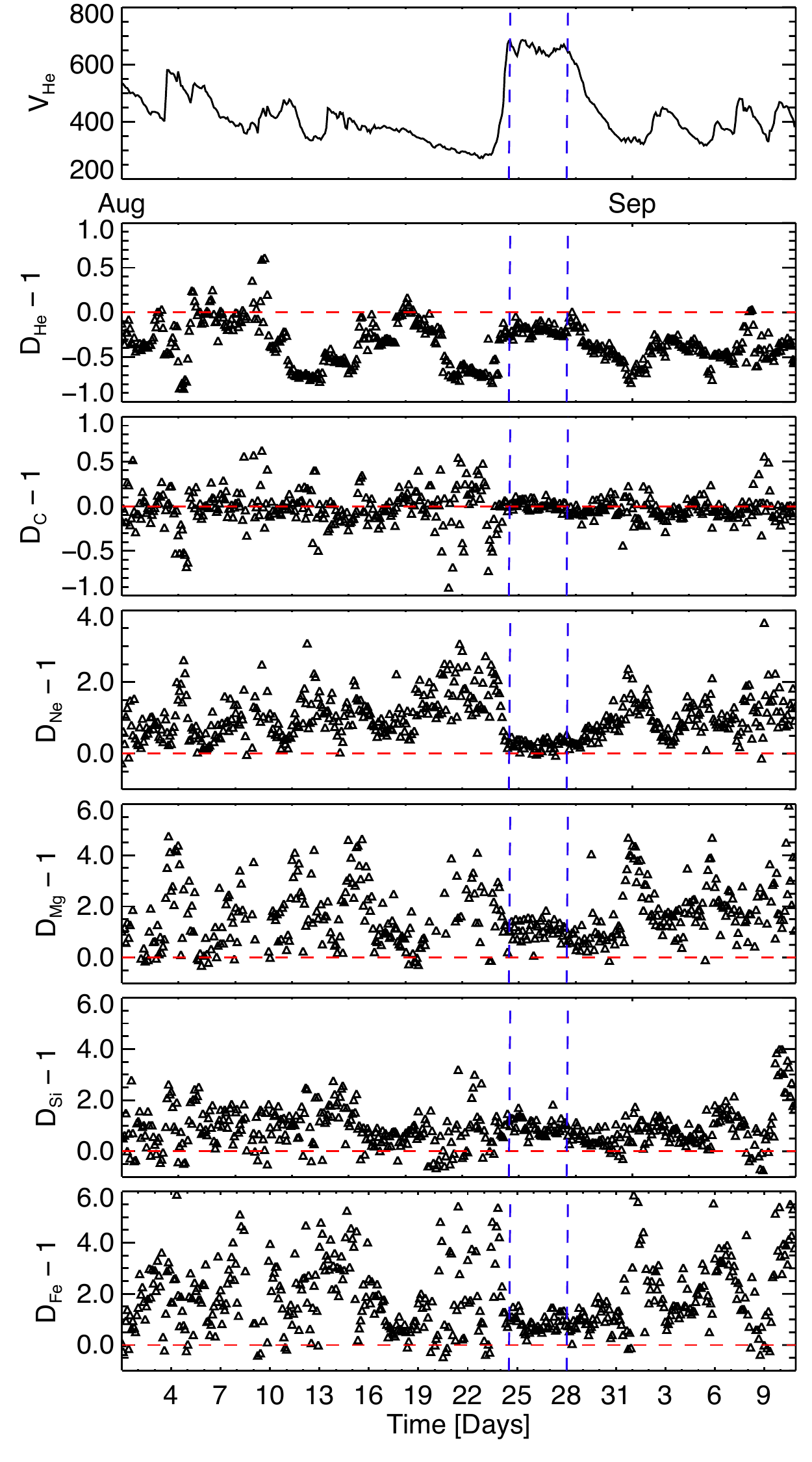}
\caption{Variation in the ``degree of fractionation'' (\dof) of several atomic species in the epoch around the fast wind stream rooted in the August 23 2010 ECH. The red dashed line shows the value \dof -1 = 0 and the vertical blue dashed lines bracket the fast wind stream as identified from the Helium velocity (top panel).}\label{f8} 
\end{center}
\end{figure}

Consider Fig.~\pref{f8} as the compositional counterpart of Fig.~\pref{f5}. It shows the variation in the ``degree of fractionation'' of six atomic species for the time in which the ECH was studied. We define the degree of fractionation (or \dof{} for atom X) as the abundance of X measured in the solar wind relative to that of Oxygen (O) divided by the expected X/O abundance ratio in the solar photosphere. As we can see from the symbols in the lower panels of the plot, there is considerable variance in the values of \dof{} in the time leading up to, and following, the arrival of the fast solar wind stream at {\em ACE}. While {\em ACE} is in the fast wind stream (bracketed by the dashed vertical blue lines), the variation in \dof{} is {\em very} small for all atoms. Furthermore, we see that the values of \dof{} show some variance from the expected photospheric values (the red dashed line is \dof{} - 1 = 0): Helium is under-abundant by about 20$\%$, Carbon is of equivalent abundance, Neon is 4$\%$ over-abundant, and Magnesium, Silicon, and Iron are over abundant by about 50$\%$.

We can represent these values and variances pictorially by drawing an analogy with Fig.~2 of \citet{1995Sci...268.1033G} where we use the ``first ionization times'' published in Table~1 of \citet{1998SSRv...85..241G} as the abscissa values to indicate ionization timescales during the plasma heating phase\footnote{The ionization times were computed for quiet Sun conditions, a temperature of 6,000K, and density of 10$^{10}$cm$^{-3}$. The value corresponds to the inverse of the ionization rate of the atom only if it occurs from the ground state and only if the effects of recombination are small. It is not clear that the last of these is a valid approximation for the chromosphere or transition region.}. The values in this plot almost exactly duplicate those published by Geiss for a polar coronal hole fast wind stream. Clearly, the species of Iron, Magnesium, and Silicon are overabundant in the fast solar wind (relative to the photosphere) and thus the process heating (and launching) the material {\em must} intrinsically imprint this fractionation on the outflowing material, as has been suggested in the literature \citep[e.g.,][]{1998SSRv...85..241G,2010GeoRL..3722101V}. The correspondence between these values and those of \citet{1995Sci...268.1033G} would appear to indicate that the basal composition of the fast solar wind is relatively invariant, but the strength and distribution of magnetic field in the polar and equatorial coronal hole sources of the streams under consideration are likely to be approximately the same. Under the assumption that the magnetic conditions (the distribution and strength of the magnetic field) in the photosphere are important in setting the heating rate of the plasma in the lower atmosphere (probably not an unreasonable assumption), it would be interesting to systematically study the change in the mixture of these atomic species in fast wind streams with the variance in the underlying photospheric magnetic field measurements \citep[e.g.,][]{2002GeoRL..29i..66Z}. Of course, in this paradigm, subtle changes in the underlying magnetism of the quiet solar atmosphere would produce subtle changes in the output charge state distributions, and possibly also the detected abundance ratios \citep[e.g.,][]{2011ApJ...740L..23M}. 

\begin{figure}
\begin{center}
\includegraphics[width=85mm]{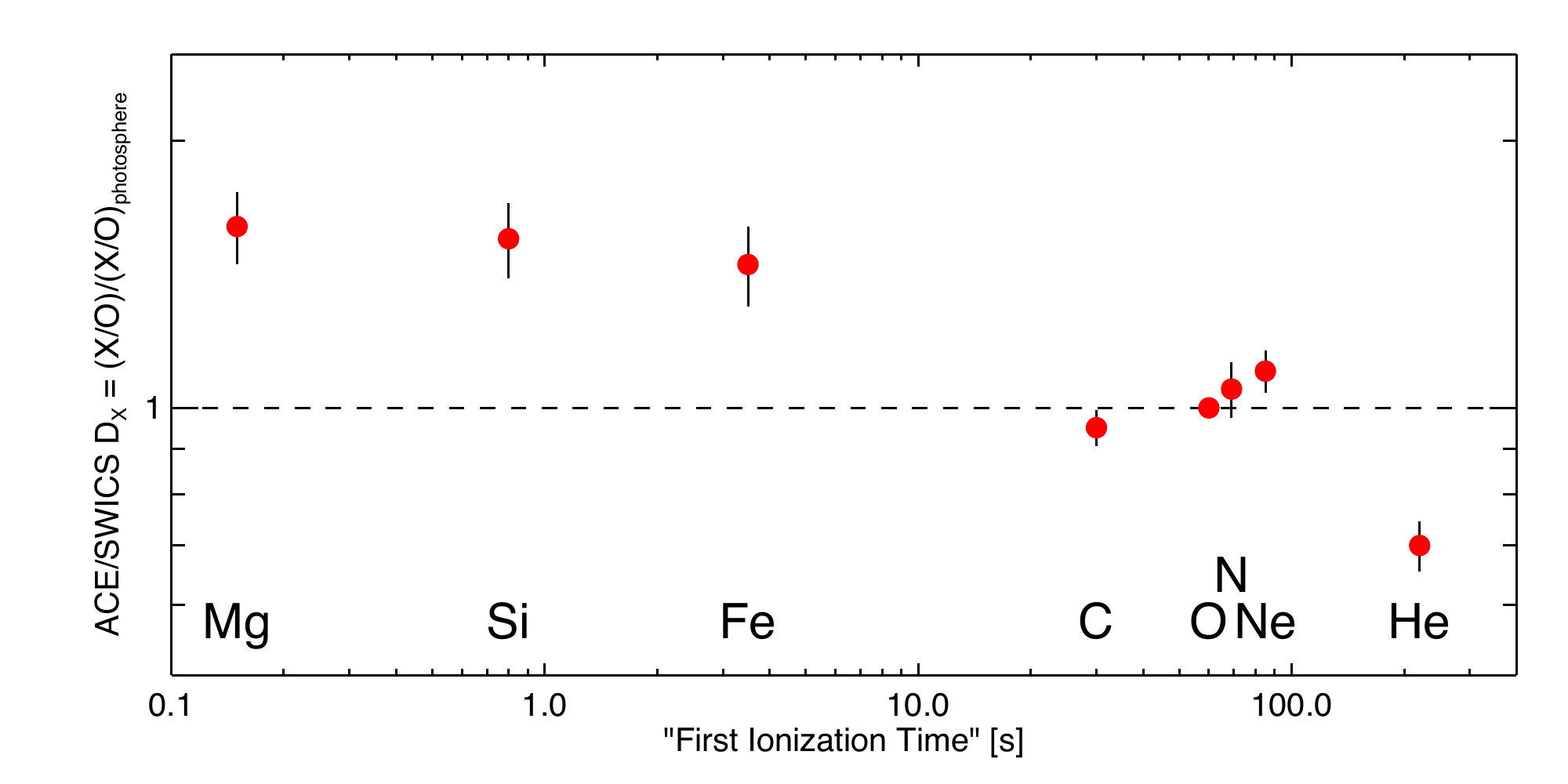}
\caption{Comparing the \dof{} computed from {\em ACE} SWICS measurements with the ``First Ionization Time'' of \citet{1998SSRv...85..241G} for the August 23, 2010 ECH.}\label{f9} 
\end{center}
\end{figure}

\section{Discussion}
Using the high S/N, low scattered light observations of AIA we can study the spatio-temporal variation of the PCDs that permeate coronal holes. At all wavelengths, but particularly those of the cooler plasmas (304 and 171\AA), we see that the coronal hole is highly dynamic and displays a great deal of small-scale, rapidly evolving, fine structure. The apparent propagation speeds of the PCDs were approximately consistent across a wide range of temperatures. More complex, lower S/N, spectroscopic observations would appear to support the hypothesis that the coronal hole plasma is ``flash'' heated at the bottom (to at least 2MK; the equilibrium peak formation temperature of \ion{Fe}{14} the dominates the AIA 211\AA{} channel), the material is then ejected outward/upward (visibly emitting to at least 1MK - \ion{Fe}{9}), consistent with the observations of mass-motion and heating associated with Type-II spicules \citep[][]{2011Sci...331...55D}. The PCDs have typical lifetimes ranging from 50 to 200s. Assuming 10$^{9}$ cm$^{-3}$ for a Type-II spicule density, a duration of 100 seconds, a velocity of 50km/s, a cross-section of 100km, we place an lower estimate (based solely of the kinetic energy of the material) required per PCD as approximately 10$^{24}$~erg, a value clearly dependent on assumptions concerning the actual PCD parameters, but is very much in the ``nanoflare'' range \citep[e.g.,][]{1988ApJ...330..474P}. Subsequent investigations will study the physical process (or processes) that drive the spicules and PCDs, heating a significant fraction of their mass to temperatures greater than 1~MK. It is likely that plasma experiments \citep[e.g.,][]{2010AIPC.1242..156B} and computational simulations which show the presence of similar PCD velocities \citep[e.g.,][]{1986SoPh..103..299S, 2011ApJ...736....9M} will provide essential clues.

A detailed study of the movies supporting Fig.~\pref{f5} would appear to indicate that there are at two scales of PCDs visible in coronal holes. On the larger, supergranular scale there are strong outflow concentrations like those discussed in the previous sections. Those are visible across the AIA channels and are rooted in regions where the magnetic field concentration is of the same polarity as the ECH (negative). These PCDs are also in very close proximity to one another (e.g., Fig.~\pref{f6}). This, perhaps, suggests that objects like quasi-separatrix-layers \citep[or QSLs, see e.g., Sect. 5.2 of][and references therein]{2005LRSP....2....7L} are critical in understanding the energy release of the spicules and PCDs. On finer spatial scales there are weakly emitting PCDs sprinkled throughout the ECH \--- these are readily visible in the accompanying movies. This may explain our earlier discussion of Figs.~\pref{f1} and~\pref{f2}, where we noted that the blue Doppler shifts and R-B asymmetries didn't map only to the network vertices. We anticipate that comparison of this second PCD population in the high S/N 171\AA{} channel and HMI photospheric magnetograms will reveal that they are formed in, or immediately around, small emerging magnetic flux concentrations in the coronal hole.

Follow-on studies will be crucial to determine the potential impact of these heating events on the composition and speed of the fast solar wind. Numerous challenges to our understanding arise as a result of the presented material. Clearly, the plasma in the PCDs can reach temperatures in excess of 1MK, and this heated portion of the material can overcome the Sun's gravitational pull to become a constituent of the solar wind. Is that heating fingerprint ``encoded'' directly in the charge state and elemental distributions measured at 1~AU? Furthermore, it is not clear that this initial ``push'' is sufficient to maintain the 650km/s speed of the fast wind stream. So, is there an additional energy source associated with these PCD structures, one that can continuously drive the material out into the heliosphere? There are strong low-frequency \alfvenic{} motions associated with the PCDs \citep{2007Sci...318.1574D, 2011Natur.475..477M}, that according to recent models, can provide the energy needed to accelerate the plasma to typical fast wind speeds and maintain it \citep[e.g.,][]{1991ApJ...372..719P, 2005ApJS..156..265C, 2005ApJ...632L..49S,2006ApJ...640L..75S,2007ApJS..171..520C,2007ApJS..171..520C,2010ApJ...708L.116V}. We speculate that the fast solar wind captured at 1AU is the result of a two-stage process, the first stage heats the plasma, and then the second drives the outward acceleration to the measured speeds. In that case, the wave spectrum of the  second stage will be critical in establishing the final speed of that wind \citep[][]{2007ApJS..171..520C}, an deduction that is observationally testable with instruments like the Coronal Multi-channel Polarimeter \cite[CoMP;][]{Tomczyk2007}. 

This two-stage paradigm is also consistent with studies of the fast wind composition \citep[e.g.,][]{1995Sci...268.1033G} where the plasma heating must occur in the lower solar atmosphere \citep[][]{1998SSRv...85..241G} to produce the mixture of abundances and charge-states observed. We expect that the mixture of abundances and charge-states prevalently seen in the fast solar wind can be used to constrain the strength and duration of the heating process taking place at the source. However, while the exact details of the heating process and their relationship to the underlying magnetic field remain unknown, the upcoming launch of the Interface Region Imaging Spectrograph (IRIS) will provide additional high spatio-spectoral-temporal resolution spectroscopic observations that will allow us to directly compare with imaging observations and test the two-stage hypotheses presented above.

\section{Conclusions}
We have discussed the rapid plasma motions and apparent heating that are visible in the magnetic network of coronal holes \--- it is clear that the instruments on \sdo{} offer a fantastic opportunity for the community to study the basal mass and energy release into the heliosphere. Some of the plasma in the dynamic heating events at the base of the fast solar wind is heated to temperatures of at least 1~MK, leaving the surface at speeds of the order of 100km/s. The process that establishes and maintains the high speed solar wind streams measured at 1AU requires subsequent investigation, but we note that a significant flux of \alfvenic{} wave energy is present on the same structures. Advances in solar wind modeling will be required to accommodate the observed duality of quasi-periodic heating and mass release coupled to the propagation, and necessary dissipation, of the abundant \alfvenic{} motions.

\begin{acknowledgements}
I would like to thank THZ for encouraging me to write this paper reviewing our recent observational investigations, Bob Leamon, Hui Tian, and Bart De Pontieu for frequently (and vigorously) discussing these issues. Thanks also to Laurel Rachmeler for helping me parse some of the text. I apologize for not throughly reviewing the vast coronal hole spectroscopy literature - hopefully the references included provide an adequate cross-section of the excellent work done in this area. The material presented was supported by the National Aeronautics and Space Administration under grant NNX08AU30G issued by the Living with a Star Targeted Research \& Technology Program. In addition, part of the work is supported by NASA grants NNX08AL22G, NNX08BA99G and NNX08AH45G and ATM-0925177 from the National Science Foundation. We are grateful that the SWEPAM team make their data available through the {\em ACE} Science Center. The National Center for Atmospheric Research is sponsored by the National Science Foundation.
\end{acknowledgements}


\begin{thebibliography}{69}
\bibitem[\protect\citeauthoryear{{Peter}}{2001}]{2001A&A...374.1108P}
\begin{barticle}
\bauthor{\binits{H.} \bsnm{{Peter}}},
\batitle{{On the nature of the transition region from the chromosphere to the
  corona of the Sun}}.
\bjtitle{\aap}
\bvolume{374},
\bfpage{1108}--\blpage{1120}
(\byear{2001}).
doi:\doiurl{10.1051/0004-6361:20010697}
\end{barticle}
\endbibitem
\end{thebibliography}

\end{document}